\documentclass[conference]{IEEEtran}
\IEEEoverridecommandlockouts
\usepackage{cite}
\usepackage{url}
\usepackage{wrapfig}
\usepackage{amsmath,amssymb,amsfonts}
\usepackage{algorithmic}
\usepackage{graphicx}
\usepackage{subcaption}
\usepackage{textcomp}
\usepackage{xcolor}
\usepackage{booktabs}
\def\BibTeX{{\rm B\kern-.05em{\sc i\kern-.025em b}\kern-.08em
    T\kern-.1667em\lower.7ex\hbox{E}\kern-.125emX}}
\begin{document}



\title{HydroTrack: Spectroscopic Analysis Prototype Enabling Real-Time Hydration Monitoring in Wearables
}

\author{\IEEEauthorblockN{Nazim A.Belabbaci$^1$, Mohammad Arif Ul Alam$^{1,2,3}$}
\IEEEauthorblockA{
\textit{$^{1,2}$University of Massachusetts Lowell}\\
\textit{$^{2}$University of Massachusetts Chan Medical School}\\
\textit{$^{3}$National Institute on Aging, National Institute of Health}\\
nazimahmed\_belabbaci@student.uml.edu, mohammadariful\_alam@uml.edu}
}

\maketitle

\begin{abstract}




In the rapidly growing field of wearable technology, optical devices are emerging as a significant innovation, offering non-invasive methods for analyzing skin and underlying tissue properties. Despite their promise, progress has been slowed by a lack of specialized prototypes and advanced analysis techniques. Addressing this gap, our study introduces, \emph{HydroTrack}, an 18-channel spectroscopy sensor, ingeniously embedded in a smartwatch.
Accompanying this hardware, we present signal processing and data analysis techniques implemented at the edge, designed to maximize the utility of our system in comprehensive health tracking. A pivotal application of our device is the real-time assessment of hydration levels in physically active individuals. We validated our prototype and analytical approach through experiments on six participants, focusing on hydration dynamics during physical exercises. Our findings reveal an accuracy of avg. 95\% in determining hydration states.
\end{abstract}


\section{Introduction}


The ability to accurately monitor hydration is crucial, especially in physically demanding environments, as dehydration can impair cognitive and physical performance \cite{b1}. The challenge in hydration monitoring is finding a real-time, non-invasive method, tailored for vulnerable groups like athletes, military personnel, hospitalized patients, infants, the elderly, and those fasting for religious or cultural reasons\cite{hydration3}. Traditional hydration assessment methods like blood osmolality tests and saliva analysis are either invasive or not suitable for real-time monitoring \cite{hydration1}. This underscores the demand for innovative, user-friendly systems capable of providing accurate, real-time hydration status in various physical conditions and environments \cite{b2}. Recent developments in wearable technology offer a promising avenue for addressing these challenges\cite{b3}. Various wearable, including smartwatches\cite{hydrostasis} and smart electrodes\cite{screen}, track hydration levels by measuring electrolyte changes in sweat or assessing cellular hydration using electrical or optical sensors. However, these methods are less practical for individuals who are not sweating, highlighting the need for more versatile monitoring solutions\cite{hydration2}. 


Progress in near-infrared spectroscopy (NIRS) opens new possibilities for assessing hydration by using the optical characteristics of tissues, independent of skin sweat liquid analysis. Optical methods involve shining light at a specific wavelength onto the skin to collect data. A sensor then detects the light that is reflected, absorbed, or refracted. The wavelength of light is crucial for determining how deeply it penetrates the skin when transmitting an optical signal, and that’s what mainly differentiate between the different techniques including Raman spectroscopy, NIRS and other light-based spectroscopes \cite{b4}. These technological strides have been pivotal in providing a continuous, non-invasive means to gauge hydration status, thereby having the potential to mitigate risks and improve the quality of life for various at-risk populations.

Skin analysis methods vary in their trade-offs. Non-invasive approaches offer deep insights, with LED-based systems being cost-effective but limited in information. Meanwhile, spectroscopy lab equipment provides comprehensive analyses but is costly and lacks portability. Balancing cost and effectiveness remains a challenge in selecting the optimal method for skin analysis. Hereby, we develop, \emph{HydroTrack} a non-invasive monitoring system for assessing hydration levels by seamlessly integrating an affordable spectrophotometer into an open-source smartwatch. The prototype offers on-device data analytics and evaluation, eliminating the reliance on Wi-Fi or cloud computing. 
Our key contributions:
\begin{itemize}
    \item Develop a non-invasive, spectrophotometer-integrated wearable watch prototype that incorporates both edge and cloud computing for real-time hydration monitoring.
    \item Develop a signal processing method based on Euler signal magnification to amplify changes in spectroscopy signals and employ machine learning at the edge for real-time hydration status detection.
    \item Validate the accuracy and feasibility of the proposed prototype and methods through the analysis of data gathered from six individuals in an exercise environment.
\end{itemize}

\section{Related Works}
\vspace{-.08in}
Several research employed optical technologies for monitoring dermal water content. Rockley's BioPtx\cite{rockley} utilizes an advanced silicon-photonics-based spectrophotometer to track hydration  in tissue-simulating phantoms. Another study \cite{multiwave} focuses on dermal water content measurement using four LEDs, emitting wavelengths at 940 nm, 970 nm, 1200 nm, and 1450 nm, demonstrated reliable results on porcine skin. A paper\cite{spectro} introduces a miniaturized CMOS spectrometer (650-900 nm) for continuous hydration monitoring in wearable health applications. It emphasizes accuracy and potential multi-endpoint measurement. A pilot study \cite{relatedwork}, presents a wearable hydration monitor using photonics sensors and diffuse reflectance spectroscopy for fluid status in hemodialysis patients. However, findings may not generalize to the broader population due to the specific focus on hemodialysis patients. 

\section{HydroTrack Hardware Prototype Design}
The \emph{HydroTrack} design principle focuses on designing a low-cost, general-purpose, non-invasive spectrophotometer. 

\subsection{Core Sensor Module: Triad Spectrophotometer}
\begin{figure}[htbp]
\centerline{\includegraphics[width=0.8\linewidth]{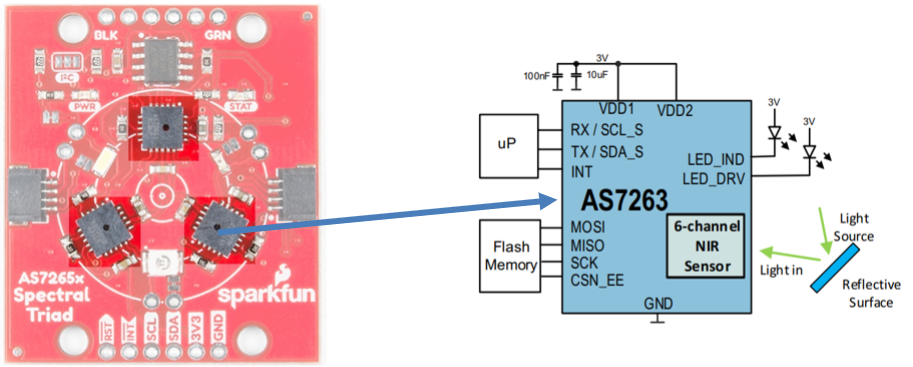}}
\caption{SparkFun Triad Spectroscopy Sensor}
\label{fig:triad}
\vspace{-.1in}
\end{figure}

We integrated a multi-wavelength spectroscopy sensor, the Triad Spectroscopy Sensor\cite{b5}, which merges the precision of clinical spectroscopy setups with the portability and affordability of LEDs. The SparkFun Triad Spectroscopy Sensor, based on three AS7265x spectral sensors, uses visible, UV, and IR LEDs to illuminate surfaces and measure their absorbance. This sensor combines three Integrated Circuits (ICs) and can detect light from 410 nm (UV) to 940 nm (IR). It is capable of measuring 18 individual light frequencies with a precision of 28.6 nW/cm2 and an accuracy of +/-12\%, thereby making high-cost equipment accessible for portable setups (Figure \ref{fig:triad}).

\subsection{Integration with Watch Module}
\begin{figure}[htbp]
    \centering
    \begin{subfigure}{0.35\textwidth}
        \centering
        \includegraphics[width=0.8\textwidth]{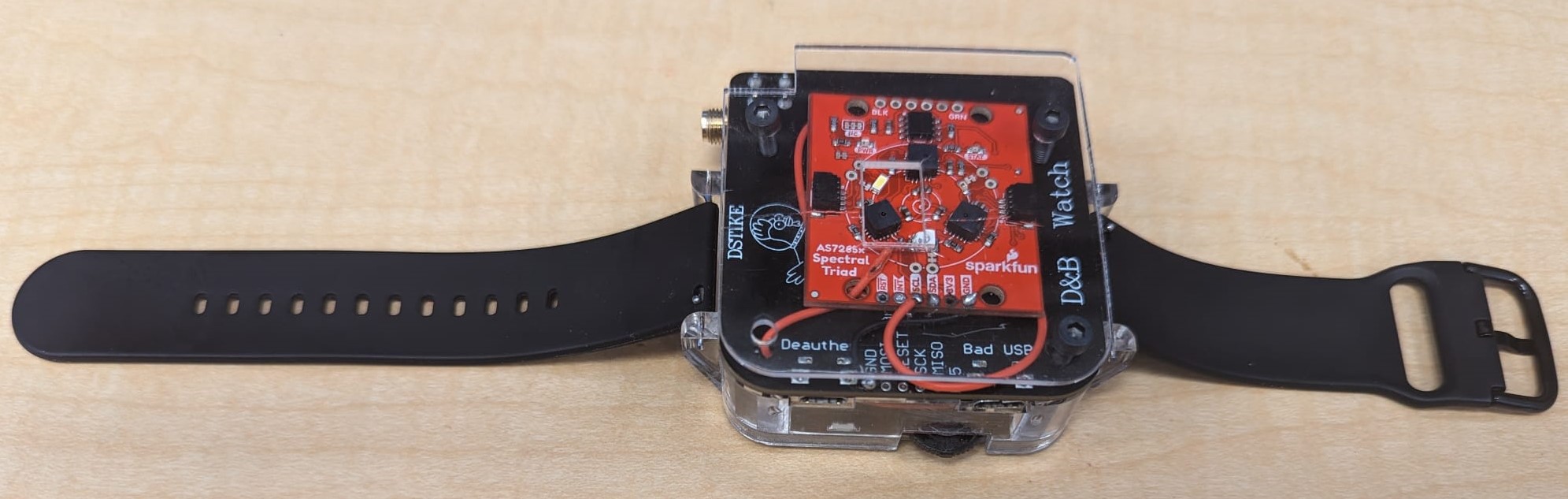}
        \label{fig:subfig1}
    \end{subfigure}
    \hfill
    \begin{subfigure}{0.35\textwidth}
        \centering
        \includegraphics[width=0.8\textwidth]{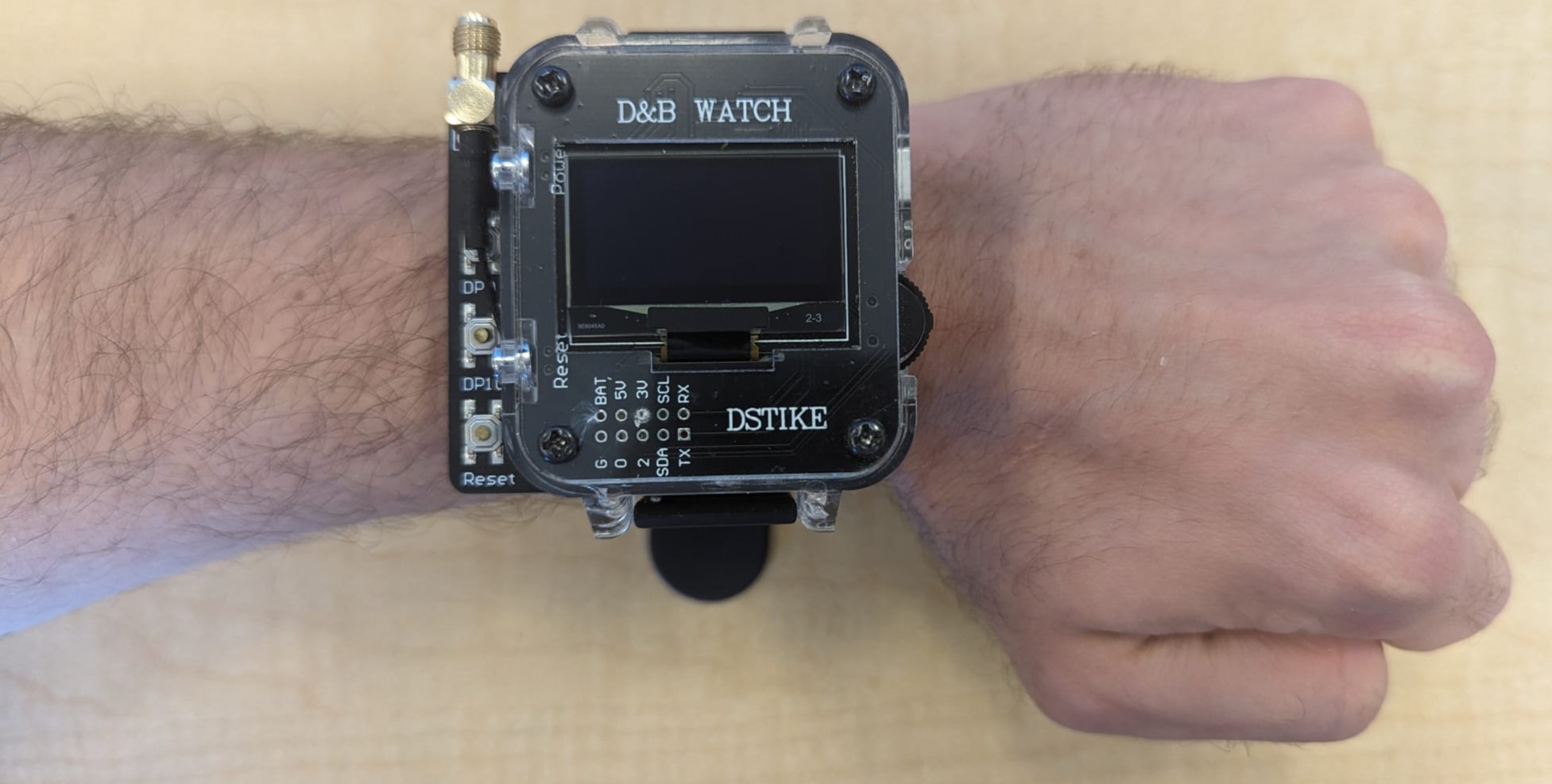}
        \label{fig:subfig2}
    \end{subfigure}
    \caption{Triad Spectroscopy Sensor integrated to DStike watch.}
    \label{fig:DStike}
    \vspace{-.1in}
\end{figure}

The DSTike Watch V4\cite{b6}, an open-source smartwatch equipped with the ESP32 micro-controller, is an ideal platform for deploying TinyML models at the edge, thanks to its favorable memory characteristics. Figure \ref{fig:DStike} illustrates the integration of the Triad spectroscopy sensor with the DSTike watch. The sensor's lights face the bottom of the watch to target the wrist. 
We used the Arduino software for programming the watch and interfacing the sensor. The final design measures approximately 80 mm x 60 mm x 22 mm (length x width x height) and weighs around 80 grams. 

The sensor measures material absorption across 18 light frequencies. In spectroscopy, Absorbance (A) is calculated as A = log(I0 / I), with I0 as the incident light intensity and I as the transmitted light intensity. I is gauged by placing the sample between the light source and sensor. This process allows for determining specific wavelength absorbance values. Calibration of the AS7265x sensor involves using a known light source as I0. 

\begin{figure*}[!htb]
\begin{minipage}{0.75\textwidth}
\vspace{-1.3in}
\begin{center}
\includegraphics[scale=0.3]{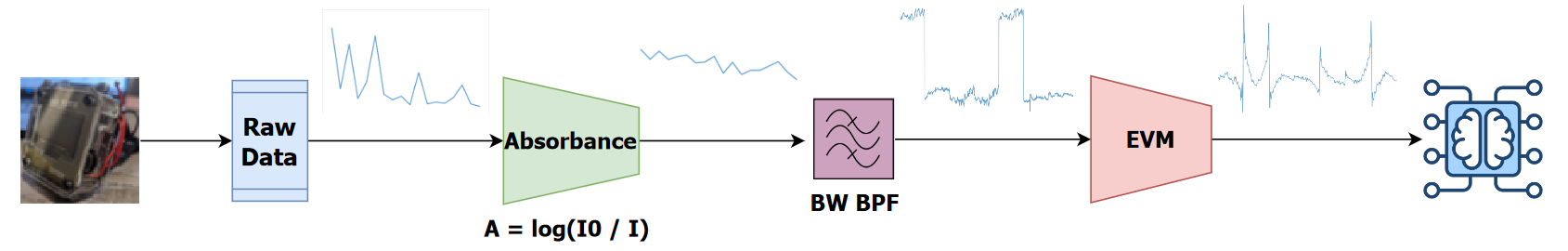}
\caption{Preprocessing pipeline of the raw spectroscopy data.}
\label{fig:process_flow}
\end{center}
\end{minipage}
\begin{minipage}{0.22\textwidth}
\begin{center}
\includegraphics[scale=0.33]{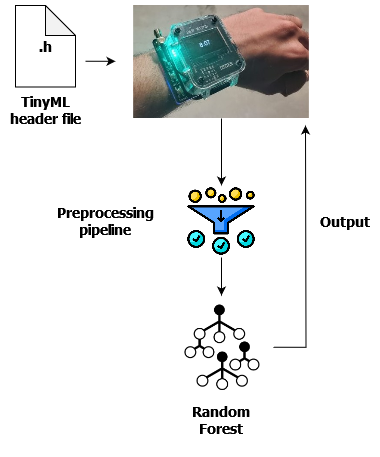}
\caption{Deploying our HydroTrack on the edge.}
\label{fig:edge}
\end{center}
\end{minipage}
\vspace{-.1in}
\end{figure*}



\section{Experiments}
\begin{figure}[!htb]
\vspace{-2in}
\begin{minipage}{0.24\textwidth}
\begin{center}

  \centerline{\includegraphics[scale=0.15]{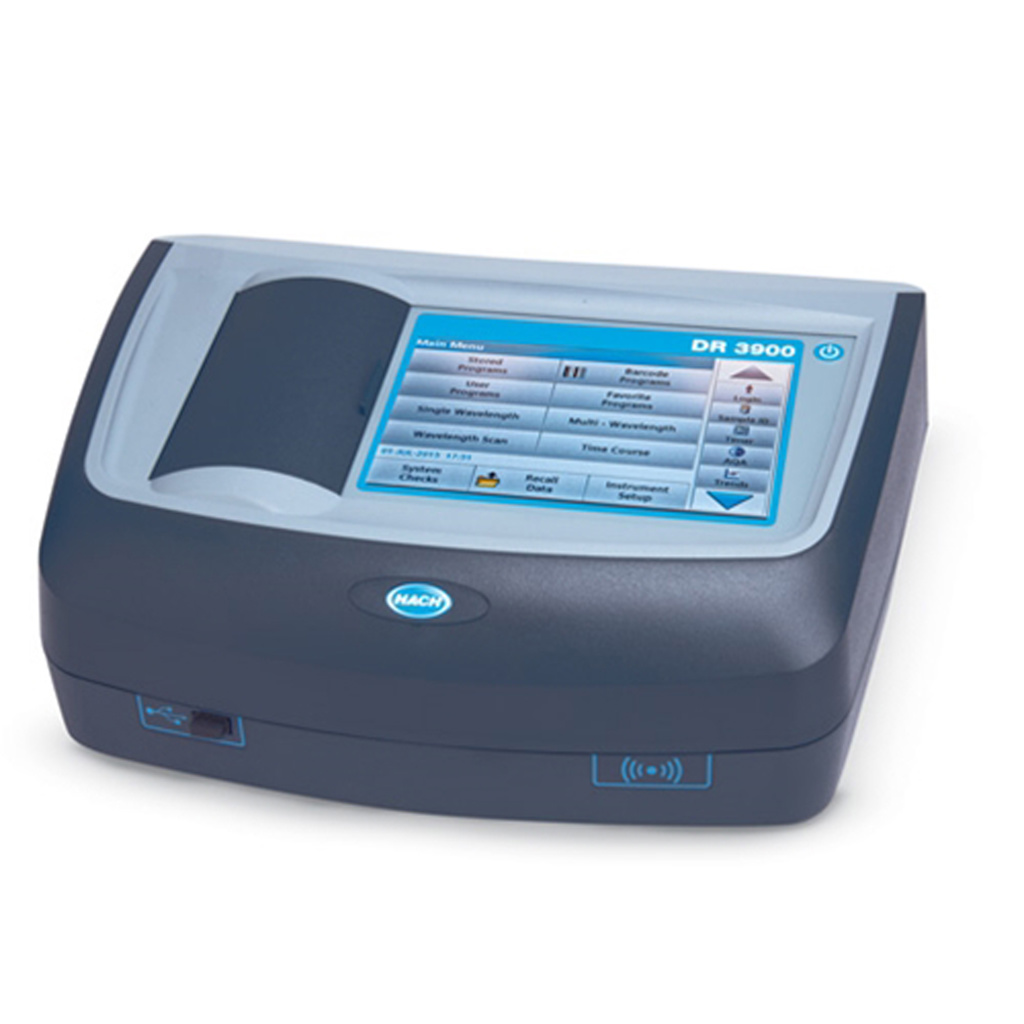}}
  \vspace{-.3in}
\caption{Hach DR3900 Laboratory Spectrophotometer \cite{b7}.}
\vspace{-.4in}
\label{fig:DR3900}
\end{center}
\end{minipage}
\begin{minipage}{0.22\textwidth}
\begin{center}
\vspace{.5in}
  \centerline{\includegraphics[scale=0.35]{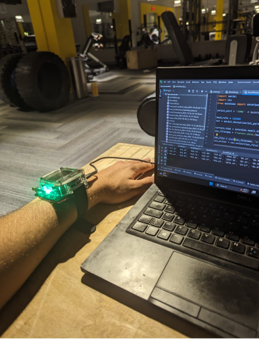}}
\caption{Hydration Tracking Experiment Setup.}
\label{fig:hydration_setup}
\end{center}
\end{minipage}
\vspace{-.3in}
\end{figure}

Before being used in real-world health monitoring scenarios, \emph{HydroTrack} must undergo validation for signal quality against a high-resolution laboratory spectrophotometer and be properly calibrated for effective real-time hydration applications. In this regard, we divided our experiment in two use cases:

{\bf Use case 1: Electrolyte solution spectroscopy:} In this experiment, we want to calibrate and benchmark our sensor with the Hach DR3900 Laboratory Spectrophotometer\cite{b7} (Figure \ref{fig:DR3900}). We opted for a liquid sample that would exhibit a distinct peak, facilitating a more straightforward comparison. For this objective, we conducted an analysis on POWERADE, a beverage renowned for containing 50\% more electrolytes than the leading sports drinks. Additionally, each bottle of POWERADE boasts 240 mg of Sodium, a crucial parameter for assessing hydration levels in blood.

{\bf Use case 2: Skin hydration monitoring}
In this scenario, our goal is to evaluate hydration levels through spectral data analysis of the skin. We conducted experiments with a cohort comprising 3 Caucasians, 2 brown skinned individuals, and one participant of Black descent, aged between 20 and 32 years. Participants were instructed to remain still and not move during data collection to avoid motion artifacts that can affect optical sensor readings. Data was collected at various time points: before the workout (well-hydrated), midway through the workout, and at the workout’s conclusion without water intake during the session. Calibration was performed by measuring the absorbance of the incident light intensity (I0) with the device off the body, in the same environment setting where the workout takes place. The process also involved using the same gains measured for every channel on the Electrolyte solution spectroscopy calibration experiment (Figure \ref{fig:hydration_setup}).

\subsection{Hydration Status Definition}
At first, we consider the data of all participants to train a model and classify 3 classes: fully-hydrated, mid-hydrated and dehydrated\cite{hydration_class}. Secondly, we isolated every participant data during the training and generate a model specific to the individual. We also conduct aggregated analyses to identify common patterns and delve into individual-level analyses to account for variations and general insights.
\subsection{Preprocessing and Feature extraction}
In our study, the absorbance data obtained during the experiment exhibited a narrow range of variation. To extract meaningful features from this limited variability, we employed Eulerian Video Magnification (EVM) \cite{evm} (Figure \ref{fig:process_flow} ). Unlike traditional methods, EVM acted as amplification technique, enhancing subtle changes within the data. This was particularly crucial in our context, as these amplified variations represent significant physiological events or anomalies that might not have been apparent in the raw, unprocessed data\cite{b8}. By utilizing EVM, we aimed to uncover nuanced patterns and fluctuations that could serve as valuable indicators of underlying physiological dynamics related to hydration status. 
We also used a Butterworth band-pass filter to reduce noise and enhance the components of interest. The signal obtained from the EVM serves as features for our hydration classifier and is used to describe the characteristics of the data. These features were fed into machine learning models to build our classifier.

\section{Results}
\subsection{Results for Use case 1}
\begin{figure*}[!htb]
\begin{minipage}{0.35\textwidth}
\begin{center}
\includegraphics[scale=0.4]{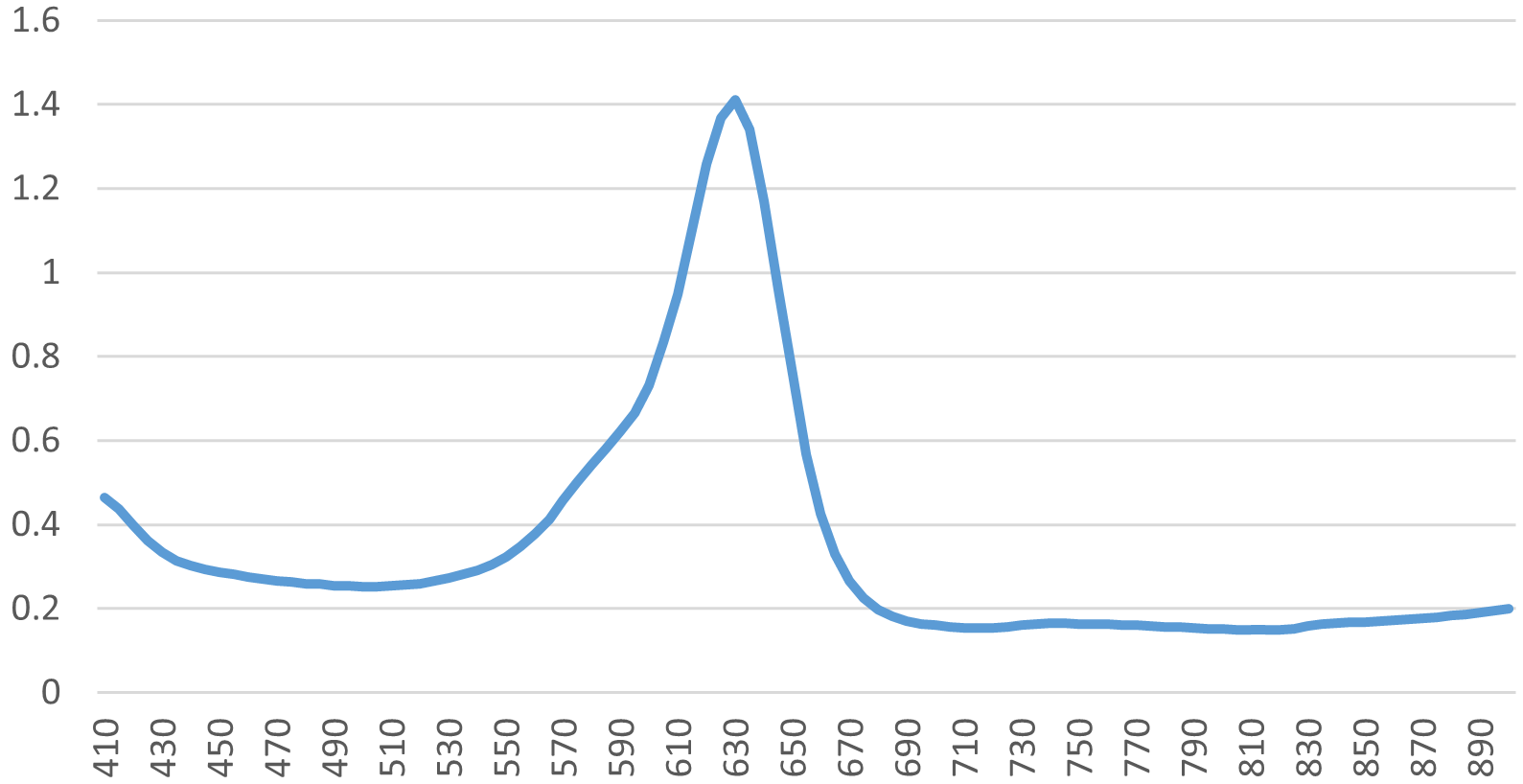}
\caption{Measuring the absorbance of the PowerADE sports drink with the DR300.}
\label{fig:result1}
\end{center}
\end{minipage}
\begin{minipage}{0.34\textwidth}
\begin{center}
\includegraphics[scale=0.4]{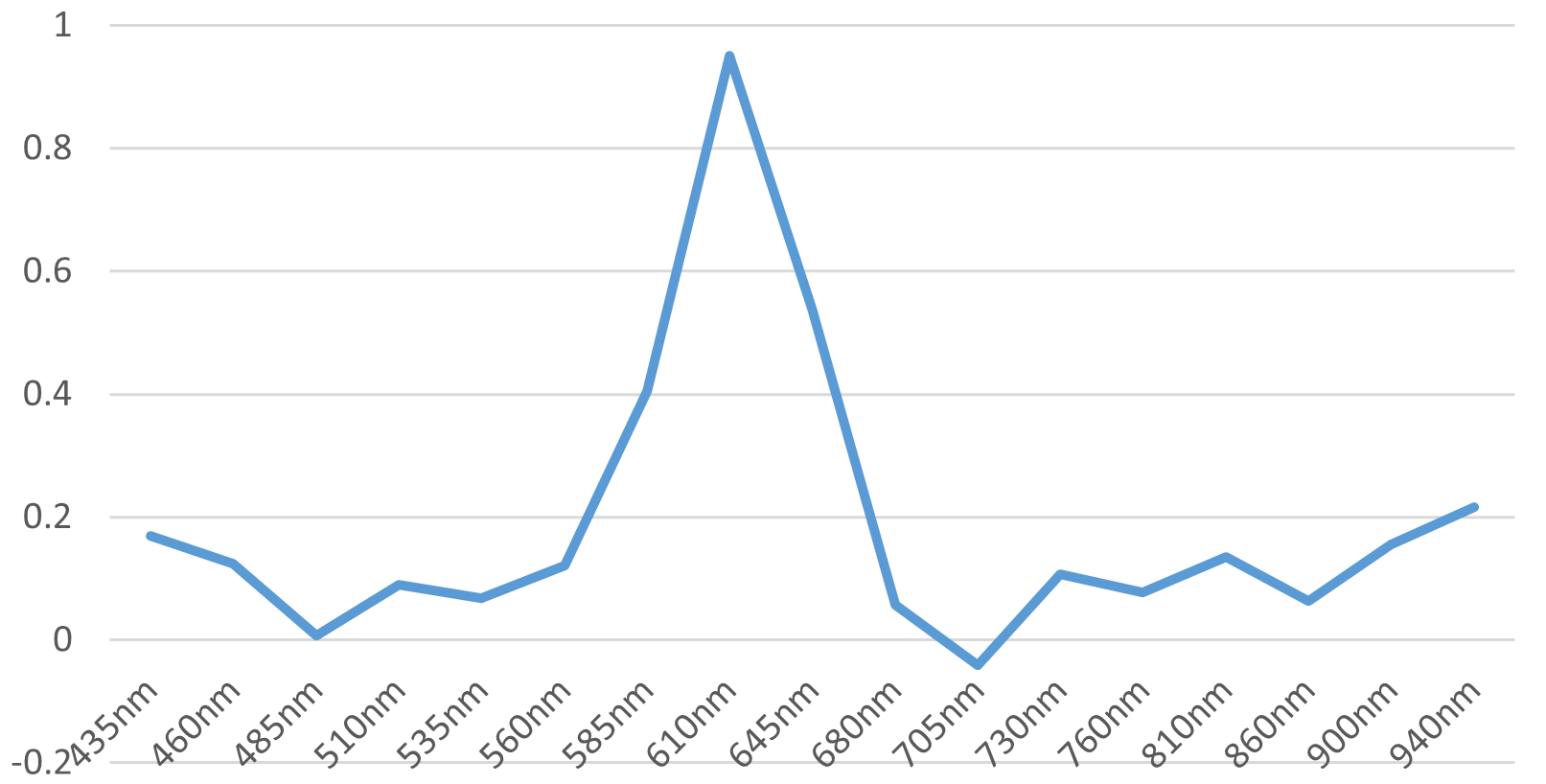}
\caption{Measuring the absorbance of the PowerADE sports drink with Triad spectroscopy sensor.}
\label{fig:result2}
\end{center}
\end{minipage}
\begin{minipage}{0.3\textwidth}
\begin{center}
\includegraphics[scale=0.4]{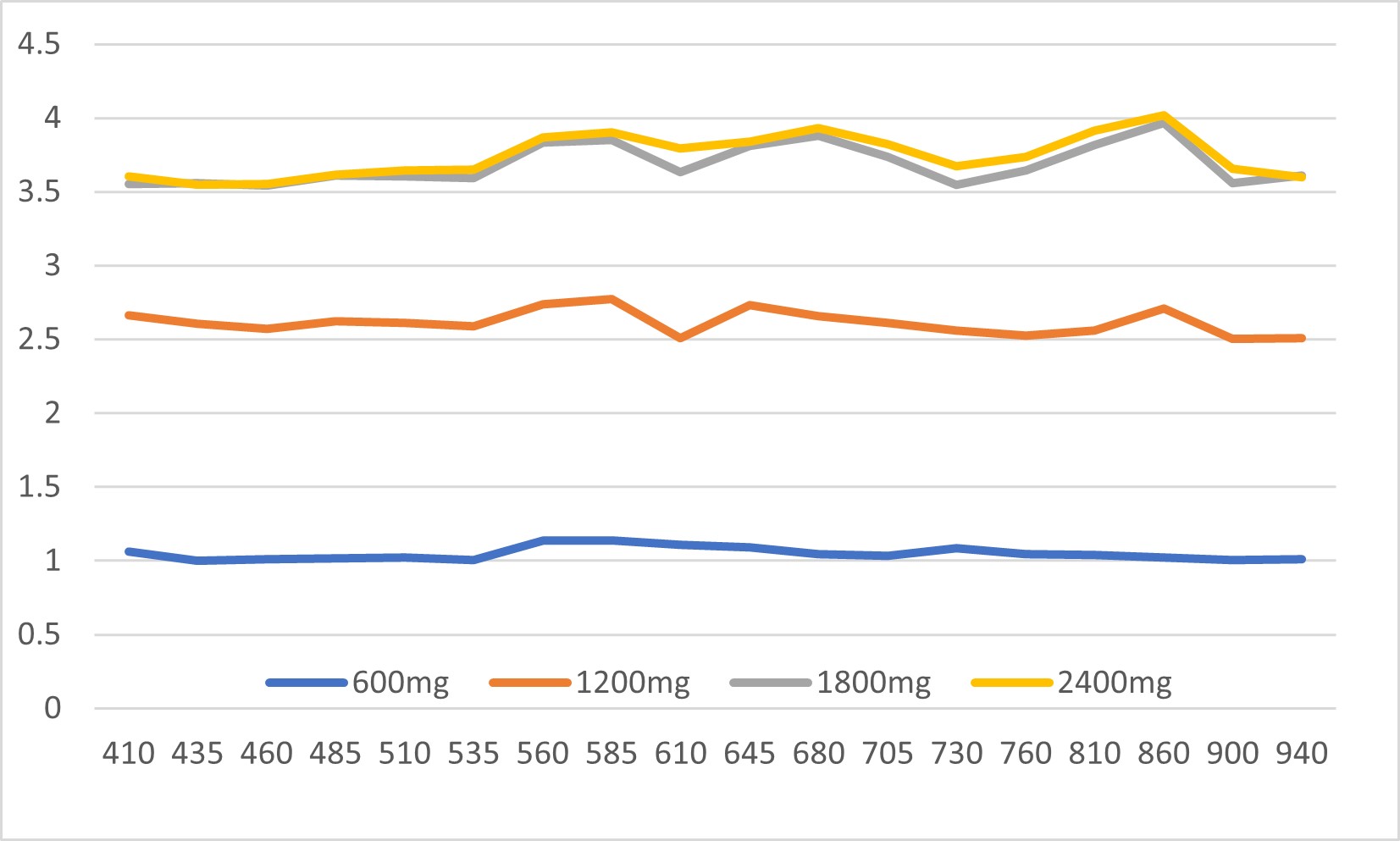}
\caption{Potassium-Chloride solution absorbance at different concentrations.}
\label{fig:result3}
\end{center}
\end{minipage}

\end{figure*}

The measurement of absorbance for the POWERADE sample with the DR3900 yielded the outcome shown in Figure \ref{fig:result1}. After calibrating the gain of some of the channels of the Triad sensor, we analysed the same sample and obtained the graph shown in Figure \ref{fig:result2}. This experiment showed us that despite the low resolution, the Triad can be reliable to further investigate other hydration biomarkers. The graphs in Figure \ref{fig:result3} is the results obtained by measuring the absorbance of a 200 mg then 400 mg sodium-chloride solution. As expected, the solution with higher concentration gives higher absorbance, with a relatively close absorbance between the results obtained using the Triad spectroscopy sensor and the DR3900.

\begin{table*}[htbp]
    \begin{minipage}{.34\linewidth}
        \centering
        \caption{Random Forest Classifier Results}
        \begin{tabular}{lcccc}
        \toprule
        Class & Prec & Rec & F1 \\
        \midrule
        Fully Hydrated & .70   &   .94   &   .80 \\
        Mid-Hydrated & .92    &   .80    &   .85   \\
        Dehydrated & .96    &  .78   &   .86 \\
        \bottomrule
        \label{tab:RF_table}
        \end{tabular}
    \end{minipage}%
    \begin{minipage}{.34\linewidth}
        \centering
        \caption{XGBoost Classifier Results}
        \begin{tabular}{lccc}
        \toprule
        Class & Prec & Rec & F1 \\
        \midrule
        Dehydrated & .96  & .95  & .95 \\
        Fully Hydrated & .93  & .96  & .94 \\
        Mid-Hydrated & .96 & .95 & .95 \\
        \bottomrule
        \label{tab:XGB_table}
        \end{tabular}
    \end{minipage}
    \begin{minipage}{.3\linewidth}
        \centering
        \caption{Accuracies for each participant: Random Forest (RF) \& XGBoost (XGB)}
\begin{tabular}{lcccccc} \toprule & P1 & P2 & P3 & P4 & P5 & P6 \\ \midrule RF & .85 & .83 & .87 & .93 & .74 & .87 \\ XGB & .94 & .89 & .91 & .95 & .86 & .89 \\ \bottomrule 
\label{tab:paticipant_table}
\end{tabular}
    \end{minipage}
    \vspace{-.3in}
\end{table*}
\vspace{-.06in}
\subsection{Results for Use Case 2}
\vspace{-.05in}
\subsubsection{Aggregate Analyses}
Aggregate analyses focusing on raw signal statistics yielded limited insights; however, using our preprocessing techniques, provided a richer understanding of the data, revealing distinct patterns representative of hydration states. Two classifiers were rigorously evaluated, with the Random Forest classifier demonstrating commendable accuracy of 84\%, and the XGBoost classifier surpassing it with an impressive 95\% accuracy. Cross-validation of the Random Forest classifier showed consistent accuracy (mean of 83\% with minimal standard deviation of 2\%). The Random Forest classifier exhibited well-learned patterns, with an average training score of 85.7\% and validation score of 82.7\%, indicating good generalization and resilience against overfitting. The XGBoost classifier outperformed in precision, recall, and F1-scores for all hydration states. XGBoost often outperforms Random Forest due to its optimized handling of gradient descent, built-in regularization techniques, and optimized tree construction. Refer to Table \ref{tab:RF_table} and  \ref{tab:XGB_table} for Random Forest and XGBoost classifier results.



\subsubsection{Individual Analysis}
Individual analyses showed varied model performance, with XGBoost consistently outperforming Random Forest. Patient-specific nuances underscore the importance of understanding individual responses to classification models. Skin pigmentation and other intra-individual variations has significant influence on optical-based sensing of human skin. Understanding these patient-specific nuances is crucial for developing effective classification models. 
Overall, both models performed well, emphasizing the pipeline success in hydration classification. The Random Forest model averaged 84\%, while XGBoost averaged 90\% accuracy. Refer to Table \ref{tab:paticipant_table} for accuracies for each participant with each classifier.



\vspace{-.07in}
\subsection{Deployment at the Edge}
We utilized TinyML with Python and the everywhereml package, choosing the Random Forest Classifier for its speed and accuracy, especially on resource-constrained devices like the DSTike watch (Fig \ref{fig:edge}). The final prototype averages absorbance data over a minute, runs the model, and displays hydration status on the OLED screen.
\subsection{Discussion}
Even though both analysis demonstrated satisfying results, the observed discrepancies between training and validation scores emphasize the need to balance complexity and generalizability for optimal classifier performance. EVM not only enhanced the training to validation ratio (from 16\% to 5\%) but also improved the overall model accuracy from 80\% to 84\%, utilizing less complex parameters. In the final model, we employed 80 estimators with a maximum depth limit of 5, whereas the original model used 100 and 10, respectively.

As mentioned in our individual results, skin color impacts light penetration. It has been demonstrated that skin color significantly influences light penetration and absorption, with absorption coefficients for dark skin being greater than light skin in the 400 to 1000 nm spectrum\cite{skincolour}.  In future work, we plan to address this challenge by Including skin color as a feature in our model and adjusting the wavelengths and intensity of light based on the skin color information. 
\section{Conclusion and Future Work}

This work introduces a novel method for real-time hydration assessment using a smartwatch with an affordable spectroscopy sensor, aiming to monitor hydration levels conveniently and non-invasively based on blood electrolyte concentration through everyday wearables. The study validated the system's reliability by exploring applications in measuring electrolyte solutions and assessing skin hydration during workouts experiments. Signal processing techniques were introduced to extract valuable data, and an AI algorithm was implemented on the edge device for efficient processing. 
While the small sample size limits generalizability, it was adequate for validating the prototype and analytical approach. Future studies will address limitations such as skin tone variations by including larger sample sizes, conducting more rigorous statistical analyses, and adjusting light intensity based on skin color information to validate findings in diverse populations. We will also explore using a motion artifact filtering algorithm to reduce motion artifact effects.

\end{document}